\documentclass[sigconf]{acmart}
\usepackage[dvipsnames]{xcolor}
\usepackage{marvosym}
\usepackage{multirow}
\usepackage{graphicx} 
\usepackage{makecell}
\usepackage{fontawesome5}
\usepackage{hyperref}
\usepackage{amsmath}

\newcommand{\topk}[1]{\mathcal{L}_k(\mathcal{#1}, q)}

\usepackage{newfloat}

\usepackage{marginnote}

\newtheorem*{axiom}{Axiom}

\usepackage[font=small]{caption}

\DeclareMathOperator*{\rel}{rel}

\AtBeginDocument{%
  }

\setcopyright{acmlicensed}
\copyrightyear{2026}
\acmYear{2026}
\setcopyright{cc}
\setcctype{by}
\acmConference[CIKM '26]{Proceedings of the 35th ACM International Conference on Information and Knowledge Management}{November 07--11, 2026}{Rome, Italy}
\acmBooktitle{Proceedings of the 35th ACM International Conference on Information and Knowledge Management (CIKM '26), November 07--11, 2026, Rome, Italy}
\acmDOI{10.1145/3799682.3839935}
\acmISBN{979-8-4007-2539-5/2026/11}

\usepackage{sigirstyle}

\begin{document}


\title{Robustness of IR Models to Collection Growth}

\author{Emmanouil Georgios Lionis}
\orcid{0009-0004-3931-9657}
\affiliation{%
  \institution{University of Glasgow}
  \city{Glasgow}
  \country{United Kingdom}
}
\email{e.lionis.1@research.gla.ac.uk}

\author{Sean MacAvaney}
\orcid{0000-0002-8914-2659}
\affiliation{%
  \institution{University of Glasgow}
  \city{Glasgow}
  \country{United Kingdom}
}
\email{Sean.MacAvaney@glasgow.ac.uk}

\author{Debasis Ganguly}
\orcid{0000-0003-0050-7138}
\affiliation{%
  \institution{University of Glasgow}
  \city{Glasgow}
  \country{United Kingdom}
}
\email{debasis.ganguly@glasgow.ac.uk}
\renewcommand{\shortauthors}{Emmanouil Georgios Lionis, Sean MacAvaney, and Debasis Ganguly}

\begin{abstract} 
Information Retrieval (IR) systems seek to identify relevant documents within a collection. In practical applications, collections are dynamic, with documents frequently added. We argue that ideally, a retriever's effectiveness should not decrease when non-relevant documents are added to a collection. This study formalises this concept and empirically evaluates it by merging two collections with negligible topic overlap. We hypothesise that the way an IR model conditions its ranking on other documents in a collection (e.g., the IDF component in BM25 or contextual documents in listwise rerankers) plays an important role in its robustness to the addition of non-relevant documents. We broadly classify models as those that do not depend on other documents (Multi-Document-Agnostic, MDA) and those that do (Multi-Document-Dependent, MDD). Our results show that neither MDD nor MDA models are fully robust to the addition of non-relevant documents, as all models exhibit some performance degradation. Interestingly, among the models we test, MDA is more effective than MDD for retrieval, whereas MDD and MDA rerankers are equally effective.

\begin{center}
\noindent
\href{https://github.com/lionisakis/subcollection}{
\faGithub \
\texttt{lionisakis/subcollection}
}
\end{center}

\end{abstract}



\maketitle

\section{Introduction}

Ad hoc retrieval, a core task in Information Retrieval (IR), aims to rank documents from a given collection according to their relevance to a user query. In real-world retrieval systems, document collections are rarely static; they continually evolve through additions, updates, and deletions as content changes over time~\cite{DBLP:conf/ecir/YangTLMMOM25, DBLP:conf/ictir/KellerBS24}.
We believe that adding new documents to a collection should not decrease retrieval effectiveness for queries whose information needs are defined with respect to the pre-existing collection. Symmetrically, retrieval effectiveness for queries targeting information needs associated with the newly added documents should remain unaffected by the presence of older, unrelated content. In summary, for any query whose relevant documents are limited to a specific subset of the collection, retrieval performance should remain invariant to the addition of documents that are non-relevant to that query. 

In the present work, we term this property as the robustness to non-relevant additions and formalise it as the collection growth axiom (\texttt{CG} Axiom). Notably, this formulation differs from prior research on out-of-domain collection robustness, 
such as multiple out-of-domain collection evaluation (BEIR~\cite{DBLP:journals/corr/abs-2104-08663}), corpus subsampling for large-scale evaluation~\cite{10.1007/978-3-031-88708-6_29}, temporal collection growth with new queries regardless of document relevance~\cite{DBLP:conf/ecir/LiuZGZRC25, DBLP:conf/ecir/YangTLMMOM25, DBLP:conf/ictir/KellerBS24}, or adversarial additions~\cite{DBLP:journals/corr/abs-2407-06992}.

\begin{figure}
    \centering
    \includegraphics[width=\linewidth]{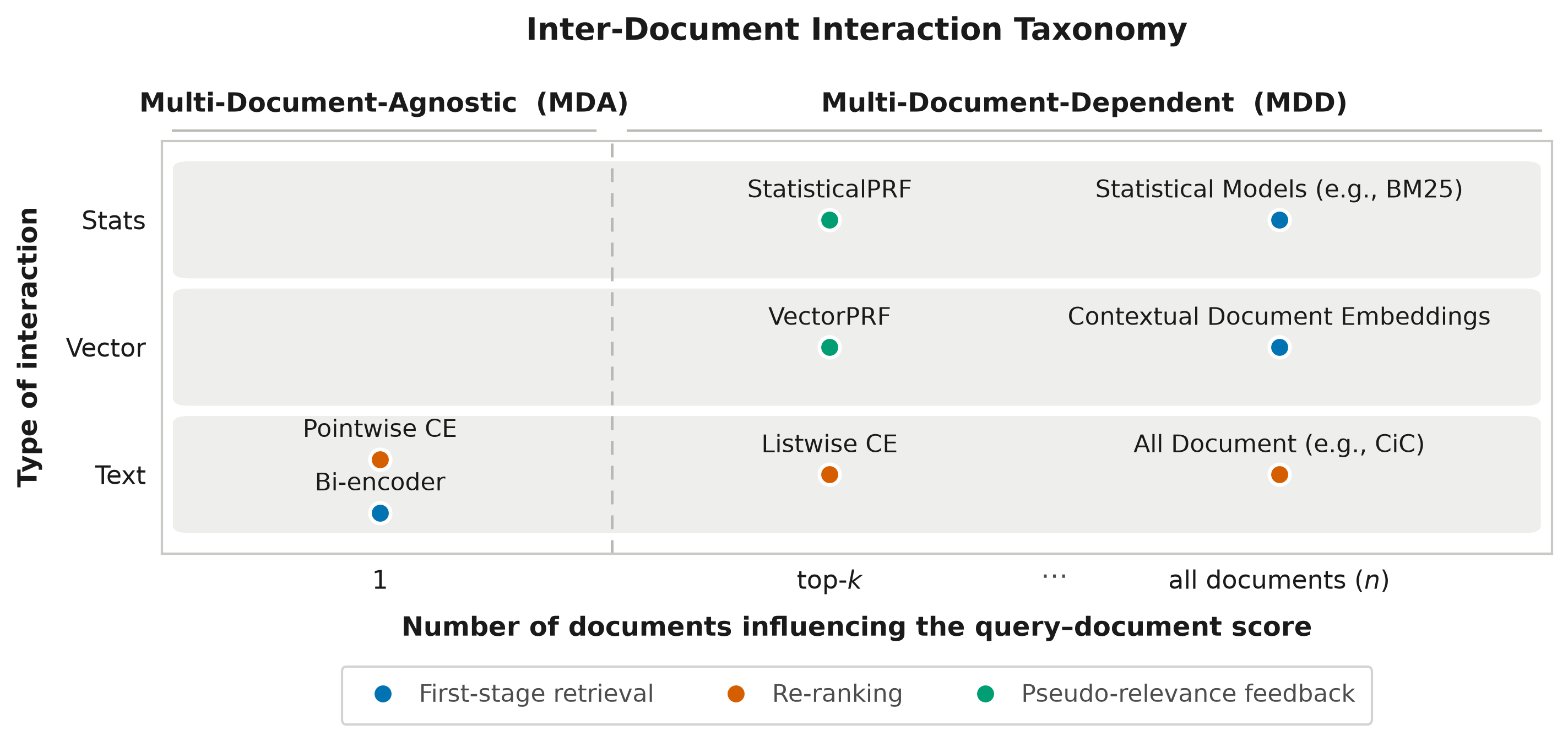}
    \caption{Inter-Document Interaction Taxonomy. Models are organised by (i) how many documents influence scoring (single, top-k, all) and (ii) interaction type (text, vector, statistics), distinguishing MDA methods that score documents independently from MDD methods that model inter-document dependencies, with representative retrieval and re-ranking approaches shown.
    }
    \label{fig:taxonomy}
\end{figure}

To empirically evaluate this robustness, we examine IR systems with multiple components that can be affected by collection growth. We consider the degree and manner in which each model accounts for other documents in the collection (their inter-document dependencies) as essential to its capacity to handle non-relevant documents within the collection.
Two principal categories of inter-document dependency are identified in this study, as depicted in Figure \ref{fig:taxonomy}.
\textbf{Multi-Document-Agnostic (MDA)} models compute relevance scores independently, without reference to other documents in the collection. Examples include Dense Retrievers (DR)~\cite{DBLP:conf/cikm/GuoFAC16,DBLP:conf/sigir/KhattabZ20}, Learned Sparse Retrievers (LSR)~\cite{DBLP:conf/sigir/FormalPC21,DBLP:conf/ecir/NguyenMY23,DBLP:conf/emnlp/NguyenCMM0Y24}, and pointwise Cross Encoders (CE)~\cite{DBLP:journals/access/ChenZZY20,DBLP:conf/emnlp/NogueiraJPL20,DBLP:conf/ecir/PradeepLZLYL22}. In contrast, \textbf{Multi-Document-Dependent (MDD)} approaches incorporate external context or collection-level information into the scoring process. Some MDD models are conditioned on the top-$k$ documents returned by a retriever, such as Pseudo-Relevance Feedback~\cite{DBLP:journals/tois/0009MZKZ23,DBLP:conf/trec/JaleelACDLLSW04, DBLP:conf/sigir/LavrenkoC01,DBLP:conf/cikm/RoyGMJ16} and listwise CE~\cite{DBLP:conf/ecir/SchlattFSZKZSPH25,DBLP:journals/corr/abs-2312-02724,DBLP:conf/emnlp/0001YMWRCYR23}. Others do not rely on explicit conditioning and instead gather contextual information from the entire collection, as in clustering-based methods~\cite{DBLP:conf/cikm/SheetritK19}, including Contextual Document Embeddings (CDE)~\cite{DBLP:conf/iclr/MorrisR25}, traditional probabilistic retrieval models~\cite{DBLP:conf/trec/RobertsonWB98,DBLP:conf/ecir/Amati06,DBLP:journals/tois/AmatiR02, DBLP:conf/sigir/GangulyRMJ15} and Corpus-in-Context (CiC)~\cite{DBLP:journals/corr/abs-2406-13121}.
Such models are typically employed in a multi-stage pipeline comprising first-stage retrievers and second-stage re-rankers~\cite{DBLP:conf/ictir/MacdonaldT20,DBLP:journals/tois/GuoCFSZC22,DBLP:conf/ecir/NguyenMY23}, where retrievers select a candidate set from a large corpus, and re-rankers refine the ranking to improve precision at top ranks.

To empirically test the robustness of an IR model to the addition of non-relevant documents, we require an evaluation setting that expands the collection while preserving relevance semantics. We therefore construct a heterogeneous collection by merging two standard IR benchmark collections, each with its own relevance judgements, which remain applicable to the combined corpus. We further require that the two collections have negligible topical overlap, so that documents from one collection can be treated as non-relevant additions with respect to queries from the other. This design provides a controlled testbed to examine whether IR models maintain retrieval effectiveness amid collection growth.

We evaluate a diverse set of IR models on this heterogeneous benchmark to assess their behaviour under collection growth. Our analysis reveals that existing retrieval models exhibit varying degrees of robustness to the addition of non-relevant documents. These observations point to a systematic limitation of current IR architectures in evolving collections and motivate the need for retrieval models explicitly designed to handle collection growth.

In summary, this work offers the following contributions:
(1) we propose the measurement of IR systems based on their robustness to the addition of non-relevant content via the mathematical \texttt{CG} Axiom;
(2) we propose a novel taxonomy of IR models characterised by types of inter-document dependency and mode of interaction query-document;
(3) We examine whether IR models with diverse architectural and scoring characteristics satisfy robustness to additional non-relevant content under controlled collection growth.

\section{Collection Growth}
\label{sec:mathematical assumption}

Collections are rarely static in production, with non-relevant documents continuously added. Thus, a well-behaved system should not demote relevant documents in response to such additions. To precisely characterise this invariance, we formalise this via an IR axiom, which states desirable system properties~\cite{DBLP:conf/sigir/FangZ06,DBLP:conf/sigir/HeinrichV0H025,DBLP:conf/sigir/SenGVJ20}. 
Let $\topk{C}$ denote the top-$k$ list of documents obtained by executing a retriever on a collection $C$. We introduce the \emph{Collection Growth (\texttt{CG}) Axiom} to formalise the characteristic of robustness to non-relevant collection expansion as a necessary rationality, as shown below:
\begin{axiom}[CG -- Collection Growth]
\label{ax:ce}
Let $\mathcal{C}$ be an existing collection of documents to which a set of new documents $\mathcal{D}$ are added to yield a larger collection $\mathcal{C}^{+} = \mathcal{C} \cup \mathcal{D}$. Retrieval performance of a model $\phi$ measured with a metric $M: \topk{C} \mapsto \mathbb{R}$ for a query $q$ should then satisfy:
\begin{equation}
\forall d \in \mathcal{D} :
\rel(q,d)=0 \Rightarrow
\frac{
M(\topk{C^+}) - M(\topk{C})
}{
M(\topk{C})
}
\leq \epsilon
\label{eq:axiomcg}
\end{equation}
\end{axiom}
\noindent where $\epsilon \in \mathbb{R}^+$ is a small positive number.
The axiom states retrieval model performance of a query whose relevant documents are limited to a specific subset of the collection ($\mathcal{C}$)
should remain invariant up to a certain small difference to the addition of documents ($\mathcal{C}^+$) that are non-relevant to that query.

One way to confirm robustness to non-relevant document addition is to verify Equation~\ref{eq:axiomcg} by measuring retrieval performance through $M$. Yet $M$ does not capture the origin of $d\in\topk{C+}$. This issue is more pronounced when $\mathcal{C}$ and $\mathcal{D}$  have different topics or information. When a query is formulated for $\mathcal{C}$, then the retriever should surface only $d\in\mathcal{C}$ as all $d\in D$ are out of topic. We quantify this with \emph{Collection Precision (CP)}, the proportion of top-$k$ results drawn from the original collection $\mathcal{C}$. Formally,
\begin{equation}
\text{CP}(\mathcal{C},\topk{C^+}) = \frac{1}{k} \sum_{d\in \topk{C^+}} \mathbb{I}\left[d \in \mathcal{C}\right].
\label{eq:CP}
\end{equation}

\section{Inter-Document Dependency}
\label{sec:inter document dependency}
This section formalises our taxonomy of multi-document dependency, showing how retrieval and ranking architectures model
inter-document dependency in relevance estimation (Figure \ref{fig:taxonomy}).

\subsection{Multi-Document-Agnostic (MDA) Models}
MDA models estimate relevance independently for each document through query--document interaction, without using inter-document or collection-level context.

\para{Bi-Encoders and Cross-Encoders (CE)}

Bi-encoders~\cite{DBLP:conf/cikm/GuoFAC16,DBLP:conf/sigir/KhattabZ20,DBLP:conf/sigir/FormalPC21,DBLP:conf/ecir/NguyenMY23, DBLP:conf/emnlp/NguyenCMM0Y24} estimate relevance by \emph{independently encoding} the query and document representations followed by a similarity score computation 
(e.g., dot-product between the embedded vectors), without conditioning on any other documents. Formally,
\begin{equation}
S_{\text{bi}}(q,d) = f(E_{\theta}(q), E_{\theta}(d)),
\label{eq:mda_biencoder}
\end{equation}
where $S_{\text{bi}}(q,d)$ denotes the estimated relevance score of document $d$ to query $q$,
$E_{\theta}(\cdot): t \mapsto \mathbb{R}^d$ represents an embedding of text $t$ into a dense vector of some dimension $d$ obtained via an encoder model $\theta$, and $f$ is a similarity function (e.g., a dot product between the embedded vectors).
In contrast to the separate encoding of queries and documents, pointwise CE~\cite{DBLP:journals/access/ChenZZY20,DBLP:conf/emnlp/NogueiraJPL20,DBLP:conf/ecir/PradeepLZLYL22} jointly encode a query and a document to estimate the relevance score. Similar to bi-encoders, in pointwise CE models the relevance computation is independent of other documents in the collection or any other collection statistics. Formally,
\begin{equation}
S_{\text{point}}(q,d) = E_\theta(q,d),
\label{eq:mda_pointwise}
\end{equation}
where similar to Equation \ref{eq:mda_biencoder}, $E_\theta$ is an encoder model.
The bi-encoders and point-wise cross-encoders are hence shown to belong to the bottom-left part of Figure \ref{fig:taxonomy}, as these both are multi-document agnostic.
Through Equations \ref{eq:mda_biencoder} and \ref{eq:mda_pointwise}, it can be seen that 
each query--document pair score is computed independently from other candidates or corpus-level information.

\subsection{Multi-Document-Dependent (MDD)}
MDD models estimate relevance conditioned on information beyond an isolated query-document pair, with scores that depend on other retrieved documents, candidate sets, or collection-level statistics, thereby inducing structured dependence through shared context, feedback signals, or the global corpus structure.

\para{Listwise Cross-Encoders}
Listwise CE
models~\cite{DBLP:conf/ecir/SchlattFSZKZSPH25,DBLP:journals/corr/abs-2312-02724,DBLP:conf/emnlp/0001YMWRCYR23} estimate the relevance conditioned on the \emph{candidate set itself} (which is often a window of documents to rerank), making each document's score dependent on the presence and content of other retrieved documents within the local window. Formally,
\begin{equation}
S_{\text{list}}(q,d)
= 
E_{\theta}(q, d, d'_1,\cdots d'_n,), 
\label{eq:mdd_listwise}
\end{equation}
where $d' \in \topk{C}$ and $n$ defined by the model ($n=1$ in a Pairwise CE~\cite{DBLP:journals/corr/abs-2101-05667} while $n=10$ in a listwise CE, like RankZephyr~\cite{DBLP:journals/corr/abs-2312-02724}).
In Equation \ref{eq:mdd_listwise} we note that in contrast to $S_{\text{point}}(q,d)$ (Eq. \ref{eq:mda_pointwise}), the similarity now depends on $\topk{C}$, which means that growing a collection from $\mathcal{C}$ to $\mathcal{C}'$ may have a more pronounced effect on the quality of the retrieved results. 

\para{Pseudo Relevance Feedback (PRF)}
PRF models~\cite{DBLP:conf/sigir/LavrenkoC01,DBLP:conf/cikm/RoyGMJ16,DBLP:journals/tois/0009MZKZ23} enrich the information need of a query by including additional terms (in sparse representations) or modifying the query vector (for dense representations). 
The modified query $q^+$ is then used to either rerank the initial results or execute a second-stage retrieval. Formally, the PRF-based scoring is of the form:
\begin{equation}
S_{\text{PRF}}(q,d) 
= f(q^+, d),\, \text{where}\, q^+ = g(q,  d'_1,\cdots d'_n,),
\label{eq:rf_score}
\end{equation}
where $g$ is a function that derives the new query representation given the $d' \in \topk{C}$. Typically $n=4$ in PRF models~\cite{DBLP:conf/sigir/LavrenkoC01,DBLP:conf/cikm/RoyGMJ16,DBLP:journals/tois/0009MZKZ23}.

\label{sec:PRF}
Equation~\ref{eq:rf_score} manifests differently under statistical and vector-based implementations, and collection growth impacts each through distinct mechanisms.
Statistical PRF~\cite{DBLP:conf/sigir/LavrenkoC01,DBLP:conf/cikm/RoyGMJ16,DBLP:conf/cikm/ZamaniDSC16} constructs $q^+$ from term co-occurrence statistics over $\topk{C}$, coupling local top-$k$ evidence with global collection statistics, a dual dependency reflected by its position in Figure~\ref{fig:taxonomy}. Collection growth therefore exposes statistical PRF through two channels: new terms $t \in \mathcal{C}^+ \setminus \mathcal{C}$ can enter $q^+$.
Dense vector PRF~\cite{colbert-prf,DBLP:journals/tois/0009MZKZ23} takes the same top-$k$ signal but encodes it as document embeddings, bypassing collection statistics entirely. Though Vector PRF may still be negatively influenced through the non-relevant documents in $\topk{C}$, as they will enrich the queries' representation.

\begin{table}[t]
\centering
\caption{Statistics and relevance label distributions for TREC-COVID and DL-2019. Hom, Het, and Het$^{+}$ represent the original, combined, and augmented qrel configurations, respectively.}
\label{tab:stats}
\small
\begin{tabular}{l rr rr r}
\toprule
\multirow{2}{*}{\textbf{Statistic}} & \multicolumn{2}{c}{\textbf{TREC-COVID}} & \multicolumn{3}{c}{\textbf{DL-2019}} \\
\cmidrule(lr){2-3} \cmidrule(lr){4-6}
& Hom & Het & Hom & Het & Het$^{+}$ \\
\midrule
Total Documents & 171K & 9.1M & 8.8M & 9.1M & 9.1M \\
Avg. Doc. Length & 197.1 & 58.3 & 56.3 & 58.3 & 58.3 \\
TREC-COVID Docs (\%) & 100.0 & 1.9 & 0.0 & 1.9 & 1.9 \\
MS MARCO Docs (\%) & 0.0 & 98.1 & 100.0 & 98.1 & 98.1 \\
\midrule
Total Queries & 200 & 200 & 50 & 50 & 50 \\
Avg. Query Length & 10.6 & 10.6 & 5.8 & 5.8 & 5.8 \\
Total Qrels & 66,336 & 66,336 & 9,260 & 9,260 & 9,273 \\
\bottomrule
\end{tabular}
\end{table}
\para{Contextual Document Encoding (CDE)}

CDE~\cite{DBLP:conf/iclr/MorrisR25} encodes document representations conditioned on local neighbourhoods of document clusters. The model induces dependency through a contextual structure derived from the entire index rather than through direct document interaction. In general, this scoring function in CDE is of the form
\begin{equation}
S_{\text{CDE}}(q,d) = f(E_{\theta}(q), E_{\theta}(d), \sum_{d' \in N_p(d)}E_{\theta}(d')),
\end{equation}
where $N_p(d)$ denotes a $p$-sized cluster of document $d$ derived during indexing. This approach relies on collection-based clusters that influence the relevance score of a query-document, thereby justifying its placement in Figure \ref{fig:taxonomy}.  Collection growth can lead to new (distractor) documents in a neighbourhood $N_p(d)$ of a document $d \in \mathcal{C}$, which, in turn, may significantly change $\sum_{d' \in N_p(d)}E_{\theta}(d')$ where $d' \in \mathcal{C}^+ - \mathcal{C}$.


\para{Lexical Retrievers}
Lexical retrievers~\cite{DBLP:conf/trec/RobertsonWB98,DBLP:conf/ecir/Amati06,DBLP:journals/tois/AmatiR02} typically use document-term weights $P(t|d)$ and corpus statistics $P(t|\mathcal{C})$ to compute similarity scores of the form
\begin{equation}
S_{\text{Lex}}(q, d) = \sum_{t \in q} f(P(t|d), P(t|\mathcal{C})),
\label{eq:mdd_bm25}
\end{equation}
and, like sparse PRF, can substantially alter scores as the collection grows due to shifts in collection statistics.

\section{Experiments and Discussion}
\label{sec:experimental setup}

\para{Research Questions} This study investigates three central Research Questions (RQ) regarding inter-document dependency in the context of robustness to non-relevant document addition collection growth:
\textbf{RQ1:} Do first-stage MDD models outperform MDA ones under collection growth?
\textbf{RQ2:} Does PRF enhance MDA retrievers by inducing MDD behaviour in a collection growth setting?
\textbf{RQ3:} Does incorporating MDD improve re-ranking performance in a collection growth setting?

\begin{figure}[t]
    \centering
    \includegraphics[width=\linewidth]{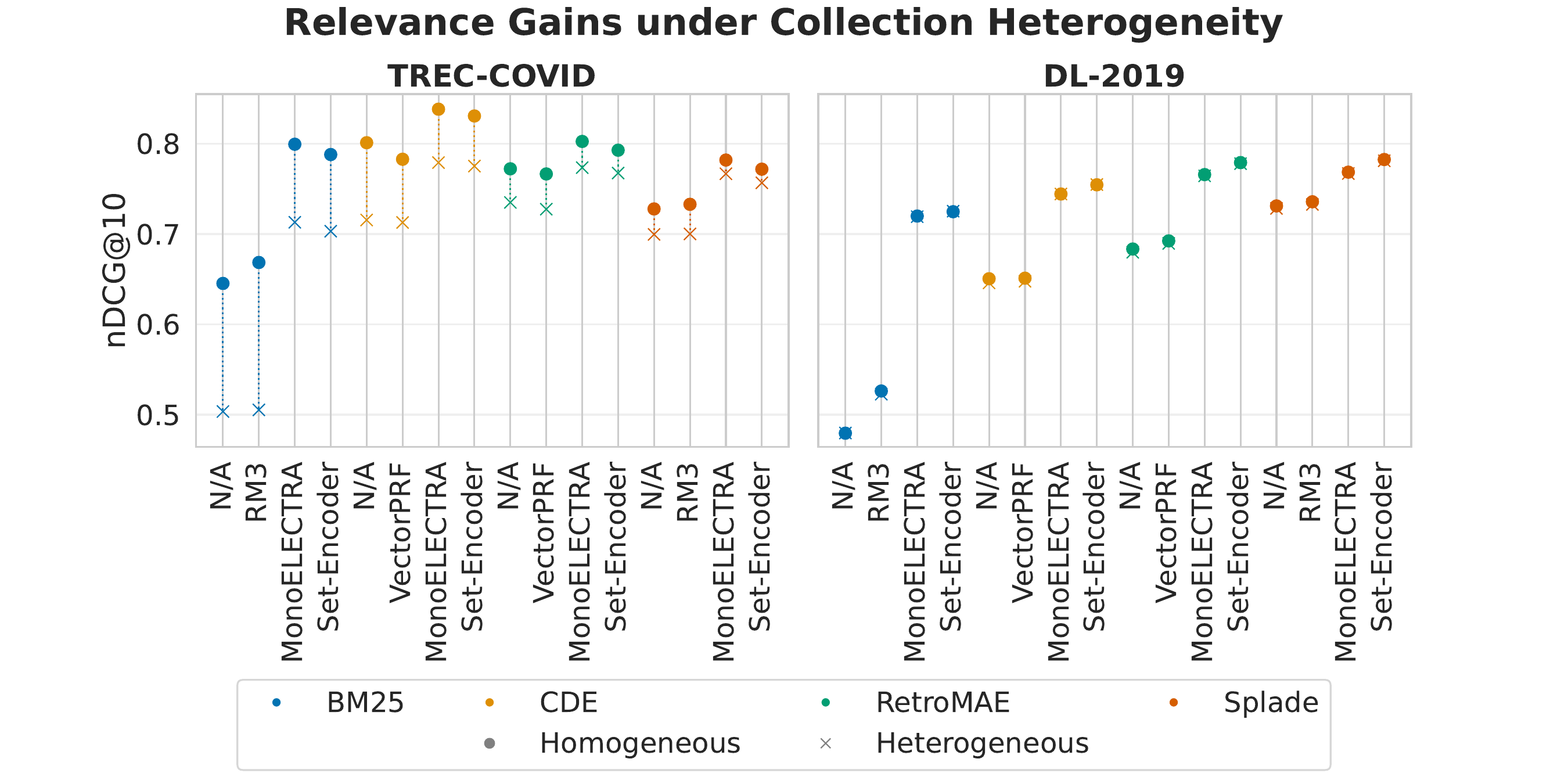}
    \caption{Relevance shifts under collection growth. Results are reported for nDCG@10 on TREC-COVID and DL-2019, evaluated on homogeneous (Circle) and heterogeneous (Cross, TREC-COVID + DL-2019) collections.}
    \label{fig:relevance_shift}
\end{figure}
\begin{table*}[t]
\caption{
Performance of different IR models under collection growth.
Results are reported for TREC-COVID and TREC DL-2019 on homogeneous (Hom) and heterogeneous (Het $\equiv$ TREC-COVID + DL-2019).
Arrows and colour indicate the sign of $\Delta(M)=(M_{\text{Het}}-M_{\text{Hom}})/M_{\text{Hom}}$.
Bold font marks the best value in each metric column, and underline indicates the best value among N/A, RM3, and VectorPRF.
Statistical significant differences ($p<0.05$) are indicated with the following: $\star$ compares CDE N/A with other N/A variants; $\diamond$ compares each Hom and Het indicated above $\Delta$; $\dagger$ compares the N/A baseline with variations in each group; $\ddagger$ compares the Set-Encoder and corresponding MonoElectra in each group.
}
\label{tab:results}
\begin{adjustbox}{width=.85\textwidth}
\begin{tabular}{@{}l@{~}l@{~~~}l@{~}c@{~~}c@{~~}c@{}c@{~~}c@{~~}c@{}c@{}c@{~~}c@{~~}c@{}c@{~~}c@{~~}c@{}c@{}}
\toprule
 &  & & \multicolumn{7}{c}{\textbf{TREC-COVID}} & \multicolumn{7}{c}{\textbf{DL-2019}} \\
\cmidrule(lr){4-10} \cmidrule(lr){11-17}
 &  & & \multicolumn{3}{c}{\textbf{nDCG@10}} & \multicolumn{3}{c}{\textbf{P@100}} & \textbf{CP@10} & \multicolumn{3}{c}{\textbf{nDCG@10}} & \multicolumn{3}{c}{\textbf{R(rel=2)@100}} & \textbf{CP@10} \\
\cmidrule(lr){4-6}
\cmidrule(lr){7-9}
\cmidrule(lr){10-10}
\cmidrule(lr){11-13}
\cmidrule(lr){14-16}
\cmidrule(lr){17-17}
\textbf{{Retriever}} & \textbf{{Reranker}} & \textbf{Type} & Hom & Het & $\Delta$ & Hom & Het & $\Delta$ & Het & Hom & Het & $\Delta$ & Hom & Het & $\Delta$ & Het \\
\midrule
\multirow[c]{4}{*}{BM25} & N/A & MDD & \phantom{0}.645$^{\tiny\textbf{$\star$}}$ & \phantom{0}.503$^{\tiny\textbf{$\star$}}$ & {$\downarrow$\phantom{0}.220$^{\tiny\textbf{$\diamond$}}$} & \phantom{0}.531$^{\tiny\textbf{$\star$}}$ & \phantom{0}.341$^{\tiny\textbf{$\star$}}$ & {$\downarrow$\phantom{0}.358$^{\tiny\textbf{$\diamond$}}$} & \phantom{0}.862$^{\tiny\textbf{$\star$}}$ & \phantom{0}.479$^{\tiny\textbf{$\star$}}$ & \phantom{0}.480$^{\tiny\textbf{$\star$}}$ & {$\uparrow$\phantom{0}\underline{.000}$^{\phantom{\tiny\textbf{$\diamond$}}}$} & \phantom{0}.488$^{\tiny\textbf{$\star$}}$ & \phantom{0}.489$^{\tiny\textbf{$\star$}}$ & {$\uparrow$\phantom{0}.002$^{\phantom{\tiny\textbf{$\diamond$}}}$} & \phantom{0}.997$^{\tiny\textbf{\phantom{$\dagger$}}}$ \\
 & RM3 & MDD& \phantom{0}.668$^{\tiny\textbf{\phantom{$\dagger$}}}$ & \phantom{0}.505$^{\tiny\textbf{\phantom{$\dagger$}}}$ & {$\downarrow$\phantom{0}.244$^{\tiny\textbf{$\diamond$}}$} & \phantom{0}.556$^{\tiny\textbf{\phantom{$\dagger$}}}$ & \phantom{0}.363$^{\tiny\textbf{\phantom{$\dagger$}}}$ & {$\downarrow$\phantom{0}.347$^{\tiny\textbf{$\diamond$}}$} & \phantom{0}.842$^{\tiny\textbf{\phantom{$\dagger$}}}$ & \phantom{0}.526$^{\tiny\textbf{$\dagger$}}$ & \phantom{0}.523$^{\tiny\textbf{$\dagger$}}$ & {$\downarrow$\phantom{0}.007$^{\phantom{\tiny\textbf{$\diamond$}}}$} & \phantom{0}.507$^{\tiny\textbf{$\dagger$}}$ & \phantom{0}.505$^{\tiny\textbf{$\dagger$}}$ & {$\downarrow$\phantom{0}.005$^{\phantom{\tiny\textbf{$\diamond$}}}$} & \phantom{0}\underline{.999}$^{\tiny\textbf{$\dagger$}}$ \\
 & MonoELECTRA & MDA& \phantom{0}.799$^{\tiny\textbf{$\dagger$}}$ & \phantom{0}.713$^{\tiny\textbf{$\dagger$}}$ & {$\downarrow$\phantom{0}.108$^{\tiny\textbf{$\diamond$}}$} & \phantom{0}.531$^{\tiny\textbf{\phantom{$\dagger$}}}$ & \phantom{0}.340$^{\tiny\textbf{\phantom{$\dagger$}}}$ & {$\downarrow$\phantom{0}.359$^{\tiny\textbf{$\diamond$}}$} & \phantom{0}.932$^{\tiny\textbf{$\dagger$}}$ & \phantom{0}.720$^{\tiny\textbf{$\dagger$}}$ & \phantom{0}.719$^{\tiny\textbf{$\dagger$}}$ & {$\downarrow$\phantom{0}.001$^{\phantom{\tiny\textbf{$\diamond$}}}$} & \phantom{0}.488$^{\tiny\textbf{\phantom{$\dagger$}}}$ & \phantom{0}.489$^{\tiny\textbf{\phantom{$\dagger$}}}$ & {$\uparrow$\phantom{0}.001$^{\phantom{\tiny\textbf{$\diamond$}}}$} & \phantom{0}.999$^{\tiny\textbf{\phantom{$\dagger$}}}$ \\
 & Set-Encoder & MDD & \phantom{0}.788$^{\tiny\textbf{$\dagger$}}$ & \phantom{0}.703$^{\tiny\textbf{$\dagger$}}$ & {$\downarrow$\phantom{0}.108$^{\tiny\textbf{$\diamond$}}$} & \phantom{0}.531$^{\tiny\textbf{\phantom{$\dagger$}}}$ & \phantom{0}.340$^{\tiny\textbf{\phantom{$\dagger$}}}$ & {$\downarrow$\phantom{0}.359$^{\tiny\textbf{$\diamond$}}$} & \phantom{0}.924$^{\tiny\textbf{$\dagger$}}$ & \phantom{0}.725$^{\tiny\textbf{$\dagger$}}$ & \phantom{0}.725$^{\tiny\textbf{$\dagger$}}$ & {$\uparrow$\phantom{-}\textbf{.001}$^{\phantom{\tiny\textbf{$\diamond$}}}$} & \phantom{0}.488$^{\tiny\textbf{\phantom{$\dagger$}}}$ & \phantom{0}.489$^{\tiny\textbf{\phantom{$\dagger$}}}$ & {$\uparrow$\phantom{0}.001$^{\phantom{\tiny\textbf{$\diamond$}}}$} & \phantom{0}.999$^{\tiny\textbf{\phantom{$\dagger$}}}$ \\
\cline{1-17}
\multirow[c]{4}{*}{CDE} & N/A & MDD & \phantom{0}\underline{.801}$^{\tiny\textbf{\phantom{$\dagger$}}}$ & \phantom{0}.716$^{\tiny\textbf{\phantom{$\dagger$}}}$ & {$\downarrow$\phantom{0}.107$^{\tiny\textbf{$\diamond$}}$} & \phantom{0}\textbf{\underline{.656}}$^{\tiny\textbf{\phantom{$\dagger$}}}$ & \phantom{0}.515$^{\tiny\textbf{\phantom{$\dagger$}}}$ & {$\downarrow$\phantom{0}.214$^{\tiny\textbf{$\diamond$}}$} & \phantom{0}.924$^{\tiny\textbf{\phantom{$\dagger$}}}$ & \phantom{0}.650$^{\tiny\textbf{\phantom{$\dagger$}}}$ & \phantom{0}.646$^{\tiny\textbf{\phantom{$\dagger$}}}$ & {$\downarrow$\phantom{0}.007$^{\phantom{\tiny\textbf{$\diamond$}}}$} & \phantom{0}.573$^{\tiny\textbf{\phantom{$\dagger$}}}$ & \phantom{0}.572$^{\tiny\textbf{\phantom{$\dagger$}}}$ & {$\downarrow$\phantom{0}.001$^{\phantom{\tiny\textbf{$\diamond$}}}$} & \phantom{0}.997$^{\tiny\textbf{\phantom{$\dagger$}}}$ \\
 & VectorPRF & MDD & \phantom{0}.783$^{\tiny\textbf{\phantom{$\dagger$}}}$ & \phantom{0}.713$^{\tiny\textbf{\phantom{$\dagger$}}}$ & {$\downarrow$\phantom{0}.089$^{\tiny\textbf{$\diamond$}}$} & \phantom{0}.628$^{\tiny\textbf{$\dagger$}}$ & \phantom{0}\textbf{\underline{.556}}$^{\tiny\textbf{$\dagger$}}$ & {$\downarrow$\phantom{0}.115$^{\tiny\textbf{$\diamond$}}$} & \phantom{0}.906$^{\tiny\textbf{\phantom{$\dagger$}}}$ & \phantom{0}.651$^{\tiny\textbf{\phantom{$\dagger$}}}$ & \phantom{0}.648$^{\tiny\textbf{\phantom{$\dagger$}}}$ & {$\downarrow$\phantom{0}.005$^{\phantom{\tiny\textbf{$\diamond$}}}$} & \phantom{0}.589$^{\tiny\textbf{\phantom{$\dagger$}}}$ & \phantom{0}.592$^{\tiny\textbf{\phantom{$\dagger$}}}$ & {$\uparrow$\phantom{-}\textbf{\underline{.004}}$^{\phantom{\tiny\textbf{$\diamond$}}}$} & \phantom{0}.996$^{\tiny\textbf{\phantom{$\dagger$}}}$ \\
 & MonoELECTRA & MDA & \phantom{0}\textbf{.838}$^{\tiny\textbf{$\dagger$}}$ & \phantom{0}\textbf{.779}$^{\tiny\textbf{$\dagger$}}$ & {$\downarrow$\phantom{0}.071$^{\tiny\textbf{$\diamond$}}$} & \phantom{0}.655$^{\tiny\textbf{\phantom{$\dagger$}}}$ & \phantom{0}.516$^{\tiny\textbf{\phantom{$\dagger$}}}$ & {$\downarrow$\phantom{0}.213$^{\tiny\textbf{$\diamond$}}$} & \phantom{0}.952$^{\tiny\textbf{$\dagger$}}$ & \phantom{0}.744$^{\tiny\textbf{$\dagger$}}$ & \phantom{0}.744$^{\tiny\textbf{$\dagger$}}$ & -\phantom{0}.000$^{\phantom{\tiny\textbf{$\diamond$}}}$ & \phantom{0}.573$^{\tiny\textbf{\phantom{$\dagger$}}}$ & \phantom{0}.572$^{\tiny\textbf{\phantom{$\dagger$}}}$ & {$\downarrow$\phantom{0}.001$^{\phantom{\tiny\textbf{$\diamond$}}}$} & \phantom{0}\textbf{.999}$^{\tiny\textbf{\phantom{$\dagger$}}}$ \\
 & Set-Encoder & MDD & \phantom{0}.831$^{\tiny\textbf{$\dagger$}}$ & \phantom{0}.775$^{\tiny\textbf{$\dagger$}}$ & {$\downarrow$\phantom{0}.067$^{\tiny\textbf{$\diamond$}}$} & \phantom{0}.655$^{\tiny\textbf{\phantom{$\dagger$}}}$ & \phantom{0}.516$^{\tiny\textbf{\phantom{$\dagger$}}}$ & {$\downarrow$\phantom{0}.213$^{\tiny\textbf{$\diamond$}}$} & \phantom{0}.952$^{\tiny\textbf{$\dagger$}}$ & \phantom{0}.754$^{\tiny\textbf{$\dagger$}}$ & \phantom{0}.755$^{\tiny\textbf{$\dagger$}}$ & {$\uparrow$\phantom{0}.000$^{\phantom{\tiny\textbf{$\diamond$}}}$} & \phantom{0}.573$^{\tiny\textbf{\phantom{$\dagger$}}}$ & \phantom{0}.572$^{\tiny\textbf{\phantom{$\dagger$}}}$ & {$\downarrow$\phantom{0}.001$^{\phantom{\tiny\textbf{$\diamond$}}}$} & \phantom{0}\textbf{.999}$^{\tiny\textbf{\phantom{$\dagger$}}}$ \\
 \cline{1-17}
\multirow[c]{4}{*}{SPLADE} & N/A & MDA & \phantom{0}.728$^{\tiny\textbf{$\star$}}$ & \phantom{0}.700$^{\tiny\textbf{$\star$}}$ & {$\downarrow$\phantom{0}\underline{.039}$^{\phantom{\tiny\textbf{$\diamond$}}}$} & \phantom{0}.562$^{\tiny\textbf{$\star$}}$ & \phantom{0}.547$^{\tiny\textbf{$\star$}}$ & {$\downarrow$\phantom{-}\textbf{\underline{.026}}$^{\phantom{\tiny\textbf{$\diamond$}}}$} & \phantom{0}\underline{.950}$^{\tiny\textbf{\phantom{$\dagger$}}}$ & \phantom{0}.731$^{\tiny\textbf{$\star$}}$ & \phantom{0}.728$^{\tiny\textbf{$\star$}}$ & {$\downarrow$\phantom{0}.004$^{\phantom{\tiny\textbf{$\diamond$}}}$} & \phantom{0}.639$^{\tiny\textbf{\phantom{$\dagger$}}}$ & \phantom{0}.635$^{\tiny\textbf{\phantom{$\dagger$}}}$ & {$\downarrow$\phantom{0}.006$^{\phantom{\tiny\textbf{$\diamond$}}}$} & \phantom{0}.997$^{\tiny\textbf{\phantom{$\dagger$}}}$ \\
 & RM3 & MDD & \phantom{0}.733$^{\tiny\textbf{\phantom{$\dagger$}}}$ & \phantom{0}.700$^{\tiny\textbf{\phantom{$\dagger$}}}$ & {$\downarrow$\phantom{0}.045$^{\phantom{\tiny\textbf{$\diamond$}}}$} & \phantom{0}.568$^{\tiny\textbf{$\dagger$}}$ & \phantom{0}.547$^{\tiny\textbf{$\dagger$}}$ & {$\downarrow$\phantom{0}.036$^{\tiny\textbf{$\diamond$}}$} & \phantom{0}.942$^{\tiny\textbf{\phantom{$\dagger$}}}$ & \phantom{0}\underline{.736}$^{\tiny\textbf{\phantom{$\dagger$}}}$ & \phantom{0}\underline{.733}$^{\tiny\textbf{\phantom{$\dagger$}}}$ & {$\downarrow$\phantom{0}.004$^{\phantom{\tiny\textbf{$\diamond$}}}$} & \phantom{0}\textbf{\underline{.643}}$^{\tiny\textbf{\phantom{$\dagger$}}}$ & \phantom{0}\textbf{\underline{.638}}$^{\tiny\textbf{\phantom{$\dagger$}}}$ & {$\downarrow$\phantom{0}.009$^{\phantom{\tiny\textbf{$\diamond$}}}$} & \phantom{0}.997$^{\tiny\textbf{\phantom{$\dagger$}}}$ \\
 & MonoELECTRA & MDA & \phantom{0}.782$^{\tiny\textbf{$\dagger$}}$ & \phantom{0}.767$^{\tiny\textbf{$\dagger$}}$ & {$\downarrow$\phantom{-}\textbf{.019}$^{\phantom{\tiny\textbf{$\diamond$}}}$} & \phantom{0}.562$^{\tiny\textbf{\phantom{$\dagger$}}}$ & \phantom{0}.547$^{\tiny\textbf{\phantom{$\dagger$}}}$ & {$\downarrow$\phantom{0}.026$^{\phantom{\tiny\textbf{$\diamond$}}}$} & \phantom{0}.966$^{\tiny\textbf{\phantom{$\dagger$}}}$ & \phantom{0}.768$^{\tiny\textbf{$\dagger$}}$ & \phantom{0}.767$^{\tiny\textbf{$\dagger$}}$ & {$\downarrow$\phantom{0}.002$^{\phantom{\tiny\textbf{$\diamond$}}}$} & \phantom{0}.639$^{\tiny\textbf{\phantom{$\dagger$}}}$ & \phantom{0}.635$^{\tiny\textbf{\phantom{$\dagger$}}}$ & {$\downarrow$\phantom{0}.006$^{\phantom{\tiny\textbf{$\diamond$}}}$} & \phantom{0}.998$^{\tiny\textbf{\phantom{$\dagger$}}}$ \\
 & Set-Encoder & MDD & \phantom{0}.772$^{\tiny\textbf{\phantom{$\dagger$}}}$ & \phantom{0}.757$^{\tiny\textbf{\phantom{$\dagger$}}}$ & {$\downarrow$\phantom{0}.020$^{\phantom{\tiny\textbf{$\diamond$}}}$} & \phantom{0}.562$^{\tiny\textbf{\phantom{$\dagger$}}}$ & \phantom{0}.547$^{\tiny\textbf{\phantom{$\dagger$}}}$ & {$\downarrow$\phantom{0}.026$^{\phantom{\tiny\textbf{$\diamond$}}}$} & \phantom{0}\textbf{.970}$^{\tiny\textbf{$\dagger$}}$ & \phantom{0}\textbf{.782}$^{\tiny\textbf{$\dagger$}}$ & \phantom{0}\textbf{.781}$^{\tiny\textbf{$\dagger$}}$ & {$\downarrow$\phantom{0}.002$^{\phantom{\tiny\textbf{$\diamond$}}}$} & \phantom{0}.639$^{\tiny\textbf{\phantom{$\dagger$}}}$ & \phantom{0}.635$^{\tiny\textbf{\phantom{$\dagger$}}}$ & {$\downarrow$\phantom{0}.006$^{\phantom{\tiny\textbf{$\diamond$}}}$} & \phantom{0}.999$^{\tiny\textbf{\phantom{$\dagger$}}}$ \\
\cline{1-17}
\multirow[c]{4}{*}{RetroMAE} & N/A & MDA & \phantom{0}.772$^{\tiny\textbf{\phantom{$\dagger$}}}$ & \phantom{0}\underline{.735}$^{\tiny\textbf{\phantom{$\dagger$}}}$ & {$\downarrow$\phantom{0}.048$^{\tiny\textbf{$\diamond$}}$} & \phantom{0}.574$^{\tiny\textbf{$\star$}}$ & \phantom{0}.550$^{\tiny\textbf{$\star$}}$ & {$\downarrow$\phantom{0}.041$^{\tiny\textbf{$\diamond$}}$} & \phantom{0}.924$^{\tiny\textbf{\phantom{$\dagger$}}}$ & \phantom{0}.683$^{\tiny\textbf{\phantom{$\dagger$}}}$ & \phantom{0}.680$^{\tiny\textbf{\phantom{$\dagger$}}}$ & {$\downarrow$\phantom{0}.005$^{\phantom{\tiny\textbf{$\diamond$}}}$} & \phantom{0}.610$^{\tiny\textbf{\phantom{$\dagger$}}}$ & \phantom{0}.609$^{\tiny\textbf{\phantom{$\dagger$}}}$ & {$\downarrow$\phantom{0}.002$^{\phantom{\tiny\textbf{$\diamond$}}}$} & \phantom{0}.998$^{\tiny\textbf{\phantom{$\dagger$}}}$ \\
 & VectorPRF & MDD & \phantom{0}.766$^{\tiny\textbf{\phantom{$\dagger$}}}$ & \phantom{0}.728$^{\tiny\textbf{\phantom{$\dagger$}}}$ & {$\downarrow$\phantom{0}.051$^{\tiny\textbf{$\diamond$}}$} & \phantom{0}.579$^{\tiny\textbf{\phantom{$\dagger$}}}$ & \phantom{0}.554$^{\tiny\textbf{\phantom{$\dagger$}}}$ & {$\downarrow$\phantom{0}.042$^{\phantom{\tiny\textbf{$\diamond$}}}$} & \phantom{0}.928$^{\tiny\textbf{\phantom{$\dagger$}}}$ & \phantom{0}.692$^{\tiny\textbf{\phantom{$\dagger$}}}$ & \phantom{0}.690$^{\tiny\textbf{\phantom{$\dagger$}}}$ & {$\downarrow$\phantom{0}.004$^{\phantom{\tiny\textbf{$\diamond$}}}$} & \phantom{0}.627$^{\tiny\textbf{\phantom{$\dagger$}}}$ & \phantom{0}.624$^{\tiny\textbf{\phantom{$\dagger$}}}$ & {$\downarrow$\phantom{0}.006$^{\phantom{\tiny\textbf{$\diamond$}}}$} & \phantom{0}\underline{.999}$^{\tiny\textbf{\phantom{$\dagger$}}}$ \\
 & MonoELECTRA & MDA & \phantom{0}.802$^{\tiny\textbf{\phantom{$\dagger$}}}$ & \phantom{0}.773$^{\tiny\textbf{\phantom{$\dagger$}}}$ & {$\downarrow$\phantom{0}.036$^{\phantom{\tiny\textbf{$\diamond$}}}$} & \phantom{0}.574$^{\tiny\textbf{\phantom{$\dagger$}}}$ & \phantom{0}.550$^{\tiny\textbf{\phantom{$\dagger$}}}$ & {$\downarrow$\phantom{0}.041$^{\tiny\textbf{$\diamond$}}$} & \phantom{0}.952$^{\tiny\textbf{\phantom{$\dagger$}}}$ & \phantom{0}.766$^{\tiny\textbf{$\dagger$}}$ & \phantom{0}.764$^{\tiny\textbf{$\dagger$}}$ & {$\downarrow$\phantom{0}.002$^{\phantom{\tiny\textbf{$\diamond$}}}$} & \phantom{0}.610$^{\tiny\textbf{\phantom{$\dagger$}}}$ & \phantom{0}.609$^{\tiny\textbf{\phantom{$\dagger$}}}$ & {$\downarrow$\phantom{0}.002$^{\phantom{\tiny\textbf{$\diamond$}}}$} & \phantom{0}\textbf{.999}$^{\tiny\textbf{\phantom{$\dagger$}}}$ \\
 & Set-Encoder & MDD
 & \phantom{0}.793$^{\tiny\textbf{\phantom{$\dagger$}}}$ & \phantom{0}.767$^{\tiny\textbf{\phantom{$\dagger$}}}$ & {$\downarrow$\phantom{0}.032$^{\phantom{\tiny\textbf{$\diamond$}}}$} & \phantom{0}.574$^{\tiny\textbf{\phantom{$\dagger$}}}$ & \phantom{0}.550$^{\tiny\textbf{\phantom{$\dagger$}}}$ & {$\downarrow$\phantom{0}.041$^{\tiny\textbf{$\diamond$}}$} & \phantom{0}.960$^{\tiny\textbf{\phantom{$\dagger$}}}$ & \phantom{0}.779$^{\tiny\textbf{$\dagger$}}$ & \phantom{0}.778$^{\tiny\textbf{$\dagger$}}$ & {$\downarrow$\phantom{0}.002$^{\phantom{\tiny\textbf{$\diamond$}}}$} & \phantom{0}.610$^{\tiny\textbf{\phantom{$\dagger$}}}$ & \phantom{0}.609$^{\tiny\textbf{\phantom{$\dagger$}}}$ & {$\downarrow$\phantom{0}.002$^{\phantom{\tiny\textbf{$\diamond$}}}$} & \phantom{0}.999$^{\tiny\textbf{\phantom{$\dagger$}}}$ \\
\bottomrule
\end{tabular}
\end{adjustbox}
\end{table*}

\para{Datasets and Data Integrity.}
MS MARCO~\cite{DBLP:conf/nips/NguyenRSGTMD16} and TREC-COVID~\cite{DBLP:journals/corr/abs-2004-10706,DBLP:journals/sigir/VoorheesABDHLRS20,DBLP:journals/corr/abs-2104-08663} were combined to simulate collection growth, yielding 9.1M documents (MS MARCO: 98.1\%; TREC-COVID: 1.9\%). We term the original collection as \textbf{Homogeneous (Hom)} while the merged collection as \textbf{Heterogeneous (Het)}.
We exclusively use the DL-2019 queries~\cite{DBLP:journals/corr/abs-2003-07820} to minimise topical overlap with TREC-COVID. This is allowed because both MS MARCO passages and DL-2019 queries predate 2019, whereas TREC-COVID documents and their queries were created after 2019 and are pandemic-focused. DL-2020 queries were excluded to avoid health- or coronavirus-related bias.

\begin{sloppypar}
To estimate the total number of new relevant documents this merged collection may introduce, we conducted a relevance-estimation validation. Firstly, we found that no MS MARCO documents are relevant to TREC-COVID queries, as pre-2019 corpus entries contain no SARS-CoV-2-specific content (e.g., \texttt{docno: 410049}). For the DL-2019 queries, we manually flagged two queries (792752, 1108939) as potentially health-related. Across all retrieval pipelines, 13 TREC-COVID documents appeared in their top-$k$ lists given those queries and we assested their relevance via the Umbrella~\cite{DBLP:journals/corr/abs-2406-06519} framework, yielding $\{Rel_0{:}5,\ Rel_1{:}2,\ Rel_2{:}6,\ Rel_3{:}0\}$\footnote{Full results: \href{https://github.com/lionisakis/subcollection/blob/main/experiments/umbrella-trec-dl-2019-trec-covid/trec_dl_2019_trec_covid/umbrella_pairs_results_only_queries.json}{subcollection/.../umbrella\_pairs\_results\_only\_queries.json}}. 
The results\footnote{The table can be found at \href{https://github.com/lionisakis/subcollection/blob/main/README.md}{subcollection/.../README.md}.} with the new Umbrella relevance distribution demonstrate the same trends as with the results DL-2019 (Table \ref{tab:results}).
\end{sloppypar}

\para{Model Configuration}
We evaluate the following: 
\uls
    \item \textbf{Retrievers}: As MDD we evaluate \textit{BM25}~\cite{DBLP:conf/trec/RobertsonWB98} (statistical) and \textit{CDE}~\cite{DBLP:conf/iclr/MorrisR25} (dense), while as MDA we evaluate \textit{RetroMAE}~\cite{DBLP:conf/emnlp/XiaoLSC22} (dense) and \textit{SPLADE}~\cite{DBLP:conf/sigir/FormalLPC22} (sparse) bi-encoders.
    \item \textbf{Converters:} As MDA to MDD converters we utilize \textit{RM3}~\cite{DBLP:conf/trec/JaleelACDLLSW04} applied to lexical retrievers (BM25, SPLADE) and \textit{VectorPRF}~\cite{DBLP:journals/tois/0009MZKZ23} applied to dense retrievers (RetroMAE, CDE). Both PRFs use the top-10 feedback documents.
    \item \textbf{Rerankers:} As MDA we evaluate \textit{monoELECTRA}~\cite{DBLP:conf/ecir/SchlattFSZKZSPH25,DBLP:conf/iclr/ClarkLLM20}, while as MDD we use \textit{Set-Encoders}~\cite{DBLP:conf/ecir/SchlattFSZKZSPH25}. Both re-rank the top 100 candidates and follow the specifications of~\cite{DBLP:conf/ecir/SchlattFSZKZSPH25}.
\ule

\begin{sloppypar}
\para{Metrics.}
Ranking effectiveness is measured using nDCG@10. The quality of document first-stage candidates is evaluated with Recall at a relevance cutoff of 2 ($R(rel=2)@100$) for DL-2019 and with Precision (P@100) for TREC-COVID, reflecting the high density of relevant documents. Collection interaction is quantified using CP@10, as defined in Eq. \eqref{eq:CP}.
\end{sloppypar}

\para{Discussion}
\label{sec:discussion}
Figure~\ref{fig:relevance_shift} shows an asymmetric effect of collection heterogeneity. 
Specifically, TREC-COVID performance $(\Delta nDCG@10)$ drops $\epsilon\approx[-0.244,-0.019]$ under heterogeneous conditions, while DL-2019 remains stable $\epsilon\approx[-0.009,+0.002]$. 
As the $\epsilon$ is large in TREC-COVID, no model satisfies the \texttt{CG} Axiom (Eq. \ref{eq:axiomcg}). 
In contrast, for DL-2019, the observed stability in performance ($\epsilon$ is very small) and near-perfect CP@10 ensure that the \texttt{CG} Axiom is satisfied.
This difference is attributed to the retrievers' pre-training on MS MARCO and the large dominant subcollection (98.1\%).

\para{RQ1} To verify whether MDD or MDA retrievers are robust, we evaluate them across the heterogeneous collections. Given Table ~\ref{tab:results},  MDA retrievers achieve the highest $nDCG_{\text{Het}}@10$ performance for both TREC-COVID (RetroMAE: $0.735$, SPLADE: $0.700$) and DL-2019 (RetroMAE: $0.680$, SPLADE: $0.728$). 
In contrast, MDD retrievers experience a substantial performance decrease ($\Delta nDCG@10$) on TREC-COVID (BM25: $-0.220$; CDE: $-0.107$) in comparison to MDA (approx. $0.1$).  For DL-2019 queries, MDD achieves fewer overall decreases in recall and performance compared to MDA approaches, while recall increases for BM25 ($\Delta R(rel=2)@100$: $0.002$). Overall, first-stage MDA retrievers exhibit the least degradation and remained robust to topic and temporal shifts, whereas MDD retrievers were more biased toward the larger collection and can improve performance under certain conditions.

\para{RQ2.} To verify whether PRF compensates for the absence of MDA characteristics in MDD retrievers, we evaluate the PRF component in both MDA and MDD retrievers in our setup. Given Table~\ref{tab:results}
and the TREC-COVID setting, for all retrievers, when a PRF component is added, we observe a higher decrease in $\Delta nDCG@10$ and/or in $nDCG@10$ compared to not adding any component (N/A) (RetroMAE-N/A: $0.735$ vs RetroMAE-VectorPRF: $0.728$). On the other hand, in DL-2019, the $nDC@10$ performance is consistently better when no component is added (N/A) (Splade-N/A: $0.728$ vs Splade-RM3: $0.733$). Notably, for CDE in Trec-COVID, not only does performance decrease, but CP@10 also drops from $0.924$ to $0.906$.
In other words, when we retrieve questions targeted at the larger collection, performance improves, whereas retrieving for the smaller collection hurts performance. Overall, PRF demonstrates bias toward the dominant collection and the documents retrieved in the top-k, regardless of the specific query intent.

\para{RQ3.} Lastly, to validate the robustness of MDA and MDD re-rankers given our collection growth, we evaluate them in our setting. On  Table~\ref{tab:results}
and in the TREC-COVID queries, MDD and MDA achieve nearly identical $\Delta nDCG@10$ values (CDE-MonoElectra: $0.779$, CDE-Set-Encoder: $0.775$), representing the best results across all combinations. This small difference ($0.002$) indicates comparable performance between MDA and MDD. A similar trend is observed with $CP@10$ (SPLADE-MonoELECTRA: $0.966$, SPLADE-Set-Encoder: $0.700$), with minimal difference between rerankers ($0.004$). For DL-2019, MDD achieves a higher nDCG@10 (SPLADE-MonoELECTRA: $0.767$, SPLADE-Set-Encoder: $0.781$) by $0.014$ points. Overall, both MDD and MDA rerankers perform competitively when non-relevant documents were added. Thus, both MDD and MDA rerankers perform competitively, confirming that rerankers are robust to collection growth of non-relevant documents. 


\section{Conclusion}
This study demonstrates that retrieval pipelines vary in their robustness to the addition of irrelevant documents. MDA retrievers are more robust during first-stage retrieval in heterogeneous, non-relevant document additions, while both MDA and MDD rerankers are equally robust at the reranking stage. Additionally, PRF modules exhibit bias in larger collections, rendering them ineffective when the introduced paradigm is detected.
These results highlight that current IR architectures exhibit systematic limitations as collections grow. Addressing this requires retrieval models that account for both pipeline stage and evolving collection characteristics from the outset.

\section*{Limitations}
Our study simulates collection growth with a single corpus pairing, MS MARCO and TREC-COVID, where the injected subcollection forms only 1.9\% of the merged corpus. However, the CG Axiom and the MDA/MDD taxonomy remain corpus-agnostic, and they extend to other domains and injection ratios in future work. This imbalance, together with the shared MS MARCO pre-training, may drive our results as much as inter-document dependency itself. 

\section*{GenAI Usage Disclosure}
Generative AI was used solely for coding, grammar and phrasing assistance. All scientific content, methodology, and conclusions are the authors' own.

\bibliographystyle{ACM-Reference-Format}
\bibliography{refs}

\end{document}